\documentclass[12pt]{article}
\usepackage{amssymb,threeparttable}

\usepackage{amsmath}
\newtheorem{la}{Lemma}[section]
\newtheorem{pro}{Proposition}[section]
\newtheorem{thm}[la]{Theorem}
\newtheorem{cor}[la]{Corollary}
\newtheorem{den}{Definition}
\newtheorem{example}{Example}

\def\real{\mathbb{R}}

\def \y {{{\mathbf y}}}

\def \s {{{\mathbf s}}}
\def \r {{{\mathbf r}}}
\def \x {{{\mathbf x}}}
\def \b {{{\mathbf b}}}

\def \v {{{\mathbf v}}}

\newcommand{\bex}{\begin{example}}
\newcommand{\eex}{\end{example}}
\def\proof{\noindent{\bf Proof.\quad}}

\def\qed{\hspace*{\fill}\vrule height6pt width4pt depth0pt\medskip}
\def\d{\displaystyle}

\title{Multi-Group Testing for Items with Real-Valued Status under Standard Arithmetic\thanks{This paper was presented in part at 2nd Japan-Taiwan Conference of Combinatorics and its Applications,
Nagoya University, Japan, 2012.}}

\author{Fei-Huang Chang$^1$, Hong-Bin Chen$^2$, Jun-Yi Guo$^3$ and Yu-Pei Huang$^4$ \\[0.3cm]
{\scriptsize $^1$Department of Mathematics and Science, National Taiwan Normal University, New Taipei, Taiwan}\\
{\scriptsize $^2$Institute of Mathematics, Academia Sinica, Taipei 10617, Taiwan}\\
{\scriptsize $^3$Department of Mathematics, National Taiwan Normal University, Taipei 11677, Taiwan}\\
{\scriptsize $^4$Department of Applied Mathematics, National Chiao Tung University, Hsinchu 30050, Taiwan}
}

\date{}
\begin{document}
\maketitle \baselineskip=24pt
\begin{center}
\bf{Abstract}
\end{center}
Motivated by applications in molecular biology and genotyping, this paper proposes a novel model of group testing for identifying items with real-valued status by using nonbinary pooling designs under standard arithmetic observation.   The purpose is to learn more information of each item to be tested rather than identify only
which ones are defectives as was done in conventional group testing. This paper provides several efficiently decodable nonadaptive
strategies for the considered problem. The major tool is a new structure called $q$-ary additive
$(w,d)$-disjunct matrix, which is related to known structures: the conventional disjunct matrix by Kautz and Singleton \cite{KS64} and the SQ-disjunct matrix by Emad and Milenkovic \cite{EM12}.


\section{Introduction}

A frequently used tool to identify an unknown set of defective (positive) elements out of a large
collection of elements by group tests is called Group Testing. In the classic group testing, a
``group'' test can be any ``subset'' of the given collection and its outcome is binary under
Boolean operations: YES or NO. The former indicates that there is a positive element in this test
and the latter implies no positive elements. Due to a diversity of its applications, there have
been many variants of the classic group testing in the literature. Readers are referred to the book
\cite{DH06} and some recent papers \cite{AFBC08,CB11,CF13,CM11,Dam05,DD11,MT11,XSTZ10} for further
information.

Most models in literature consider the elements to be tested with a binary status: $\{$positive(1),
negative(0)$\}$. In some applications, molecular biology \cite{FKKM97}, blood testing \cite{PS94}
and drug discovery \cite{XTSY01}, there can be a third category of elements called inhibitors,
anti-bodies and blockers, respectively. The presence of such an element in a test can somehow
cancel the effect of positive elements. A model addressing this issue has been intensively studied
\cite{B08,BV98,  CCH10, CCF10,CB11, DMTV01, FKKM97, HC07} under the name of group testing with
inhibitors (GTI).
Two other group testing models, mutually obscuring defectives \cite{Dam98} and
multiple access communication with interference \cite{BMTW84}, were built on the real observation
that in chemical testing and communication theory it is usually seen that there exists some
 reaction when two substances meet in a suitable condition and undesired interference when two channels receive or send a message at the same
 time.  Recently, Chen and Fu \cite{CF13} combine the above notions
and consider the multiple mutually-obscuring positives model (MMOP). In this model, more than three
categories of elements (a $k$-ary status) are allowed with an additional assumption that certain
obscuring phenomena, but unknown, occur among different categories of positive elements.


Inspired by the inhibition and the interference models, this paper considers a quantitative model
that assumes the mutual effect of inhibition and interference can be quantized through analyzing a
great amount of data in advance.  This paper focuses on the problem where the elements to be tested
are in nonbinary status and defines {\it multi-group testing}. Namely, a test can be applied to any
``multi-subset'' of the given set, where ``multi'' means every single element is allowed to be
taken more than once in a single test. The test matrix is then changed from binary in group testing
to nonbinary in multi-group testing. Note that the notion of nonbinary tests is not new and can be
found in \cite{CW99,EM13, EM12,Jevtic} with nonnegative integer matrices, \cite{CKW06} with integer
matrices and \cite{AS12} with no restriction on matrices. Obviously, allowing a number of duplicate
copies in a test is meaningless under the assumption of Boolean operations. In the considered
model, we shall assume that the outcome rule is linear under standard arithmetic.

A mathematical model can be described roughly as follows: Let $\x=(x_1,x_2,\cdots,x_n)$ $\in
\real^n$ be an unknown vector, where $x_i$ denotes the status of $i$th item. A measurement can be
applied to any vector $\y\in \{0,1,\cdots \}^n$ with an outcome $$v_{\y}\equiv f(x,y)=\langle\y,\x
\rangle=\sum_{i=1}^ny_ix_i.$$ The goal is to learn the unknown vector $\x$ through measurements in
an efficient fashion (less measurements and fast decoding).


An obvious feature under this model is that any measurement which is linearly dependent on some
other measurements is useless. The reason is that its outcome can then be simply derived from a
linear combination of the outcomes of the others. The other feature is that, without any further
information of the unknown vector $\x$, $n$ measurements are necessary in the worst case to learn
the unknown vector ($n$ is clearly sufficient). The reason is that after $k$ measurements $\x$ can
be any vector in the $n-k$ dimensional subspace whose outcome is consistent with the $k$
measurements. If $k<n$, such vectors are not unique and thus cannot be determined exactly.

To make the multi-group testing model more interesting and challenging, we shall assume that
\begin{enumerate}\item the unknown vector $\x$ is $d$-{\it sparse}, that is, $\x$
contains at most $d$ nonzero entries where $d$ is a constant with $d\ll n$; \item the entries
 of $\x$ all belong to a certain set $D\subset \real$ which is a priori knowledge;
\item the number of copies from every single item in a measurement is restricted to $\{0,1,\cdots,q-1\}$, i.e., $\y\in
\mathbb{Z}^n_q$, where the integer $q$ is prescribed. \end{enumerate} Notice that the outcome
$v_{\y}=\sum_{i=1}^ny_ix_i$ can possibly exceed $q$. Moreover, the cardinality of $D$ must be finite because the decoding algorithms
proposed in this paper rely critically on $|D|$.

This paper focuses on {\em nonadaptive strategies} where measurements are performed simultaneously
and therefore all measurements must be settled in advance. A nonadaptive strategy that uses $t$
measurements can be represented by a $t\times n$ matrix $A=[a_{ij}]$ with columns as items and rows
as measurements, and the value at $a_{ij}$ denotes the number of copies of item $j$ in measurement
$i$. The nonadaptive multi-group testing problem can then be converted into the problem: Construct
a matrix $A=[a_{ij}]$ with $a_{ij}\in \mathbb{Z}_q$ so that the  unknown $d$-sparse vector
 $\x\in D^n$ can be determined exactly and efficiently through the outcome vector $\v=A\x$.  Clearly, the conventional additive
group testing in \cite{AS85,DH00} is a special case of the multi-group testing with $D=\{0,1\}$ and
$q=2$.

\subsection*{Motivation and related work}

Our model naturally arises in several situations.  In some applications, such as blood testing,
what patients or doctors want to know might be not only a yes-or-no answer but also a more precise
index, a standard by which the level of some illness can be judged. The major purpose of relaxing
$\x$ from the usual set $\{0,1\}$ to a prescribed set $D$ (can be very large) is that we aim to
learn more information that each item carries rather than just determine which items are positive.

Of particular interest is that $D$ is allowed to contain not only positive elements but also
negative elements, whose presence is in a sense to cancel the effect of positive elements, as
inhibitors in GTI mentioned above. In GTI, the status of an inhibitor can be viewed as $-\infty$
while the outcome is still binary. Formally speaking, it is to identify an unknown sparse vector
$\x\in \{-\infty,0,1\}^n$ by measurements $\y\in \{0,1\}^n$ with two possible
outcomes:$\begin{cases}1& \mbox{ if } \langle\y,\x
\rangle \geq 1; \\
0& \mbox{ if } \langle\y,\x
\rangle <1.
\end{cases}$ However, the setting that one inhibitor is assumed to be able to cancel positive effect of all positive elements
is too powerful to be appropriate in practice. It would be more reasonable that certain weaker
cancelation effect exists between inhibitors and positives and can be quantized through analyzing a
large amount of data in advance. For instance, if the information that one inhibitor cancels $k$
positives is a priori then it can fit into the framework of our model by setting $\x\in
\{-k,0,1\}^n$.

Motivated by applications in genotyping, Emad and Milenkovic \cite{EM13, EM12} proposed the
Semi-Quantitative Group Testing (SQGT) which is a nonbinary pooling scheme combining an adder
channel and an integer-valued quantizer. The quantizer and the nonbinary $\x$ settings in SQGT and
our model, respectively, make a difference between them and no one includes the other. It is worth
mentioning that nonbinary pooling designs are used commonly. The use of nonbinary pooling designs
is based on the fact that ``genotyping methods allow for more precise readings at the output than
classical binary detectors'' \cite{EM13} and therefore the amount of samples must be reflected in
the readings. It leads to an advantage of performance, i.e., using less measurements in the
multiset model than in the set model is to be expected as set is a special case of multiset.

Recently, group testing has been related to compressed sensing in \cite{AAS13, AS12, AS13, GIS08}.
Compressed sensing is a signal processing technique for recovering a signal by finding solutions to
underdetermined linear systems (more unknown variables than equations), which coincides with the
essence of multi-group testing under standard arithmetic. As a consequence, results developed in compressed sensing could
benefit our model and vice versa. Although the two problems are in the same framework, to the best
of our knowledge, there is no research in sparse signal recovery addressing a problem with the same
setting as our model.

%
%

\subsection*{Our contribution}

We give nonadaptive strategies for the multi-group testing problem with general $D$. We note that
the one-sided case, i.e., $D$ is nonnegative or nonpositive, is a relatively simple case to handle.
The reason is that in this case a zero outcome simply implies that all the items appearing in the
measurement are zero, in contrast to the general case, a zero outcome can be produced by a
combination of some positive elements and negative elements. Although the main result for the
one-sided case has its counterpart for the general case, the technique and complexity are very
different.

We propose a new combinatorial structure called {\it $q$-ary additive $(w,d)$-disjunct matrices}
(will be defined later). Such a structure enables us to solve the general case and decode efficiently.
It is new but related to known structures: the well-known binary disjunct matrix introduced by
Kautz and Singleton \cite{KS64} and the SQ-disjunct matrix proposed by  Emad and Milenkovic
\cite{EM12}. We have a method to construct $q$-ary additive $(w,d)$-disjunct matrices, but not as
strong as we like because it relies critically on the construction of conventional disjunct matrices.
Also, we provide two methods by applying the {\it Kronecker product} to produce a bigger matrix from a smaller one.
Although the resulting matrices cannot be applied to solve the general case, they can solve the one-sided case with efficient decoding algorithms.
The
value of our constructions is not in its practicality in constructing efficient $q$-ary additive
$(w,d)$-disjunct matrices, but rather in calling awareness to the existence of such constructions,
so that further research can improve on it.

Our decoding algorithms based on $q$-ary additive $(w,d)$-disjunct matrices are quite efficient. For the one-sided case our strategy has a  decoding
algorithm in $O(|D|tn)$ time and for the general case it is $O(|D|tn^{d+1})$. By contrast, even
ignoring the time for multiplications of vectors,  in the worst case it takes extremely high time
complexity $\d O\left(\sum_{i=0}^d{n\choose i}|D|^i \right)$ to decode $\x$  by simply applying a
straightforward brute-force procedure.

The rest of the paper is organized as follows. Section 2 first  introduces notations and major
tools and then exploits them to solve the one-sided case of the multi-group testing problem under standard arithmetic. Section 3
 deals with the general case. Finally, Section 4 provides three constructions mentioned above.

\section{The one-sided case}

This section starts with a simple but useful lemma.

\begin{la}\label{linear}
Let $A$ be a matrix in $\real^{t\times n}$. Given a fixed (unknown) vector $\x\in \real^n$, let
$\v=A\x$. Let $\x'=a\x+\b$ for some $0\neq a\in \real$ and $\b\in\real^n$ (known), and let
$\v'=A\x'$. Then the problem of learning $\x$ from $\v$ is equivalent to the problem of learning
$\x'$ from $\v'$.
\end{la}
\proof The proof follows immediately from the linear mapping from $\v'$ to $\v$, that is,
$\v'=A\x'=A(a\x+\b)=aA\x+A\b=a\v+A\b$.\qed

 To present our algorithms, we first introduce some notations.
Throughout this paper, let $\x\in D^n$ be an unknown $d$-sparse vector, $A=[a_{ij}]$ of size $t\times n$ be the matrix
corresponding to the measurements and $\v=A\x$ be the outcome vector.
 For any vector $\y=(y_1,\cdots,y_n)$,
denote by $\|\y\|_0\equiv |\{y_j: y_j\neq 0\}|$ the {\it $l_0$ norm} (or {\it sparsity}) of $\y$. Given any vector $\y=(y_1,\cdots,y_t)$, for each
$j\in [n]$ define
$$t_j(\y)= |\{i:a_{ij}>y_i\}|.$$
 For any vector $\y=(y_1,y_2,\cdots,y_n)$ and $\delta\in \real$, define
$\s^{\y,=\delta}=(s^{\y,=\delta}_1,s^{\y,=\delta}_2,\cdots,s^{\y,=\delta}_n)$ where $$s^{\y,=\delta}_j=\begin{cases} 1 & \mbox{ if } y_j=\delta, \\
0 & \mbox{otherwise}.
\end{cases}$$ For convenience, we shall use $s_j$ to denote $s^{\y,=\delta}_j$ if no confusion occurs
without specified superscripts. Likewise, we define $\s^{\y,\geq\delta}$ and $\s^{\y,\leq\delta}$
by replacing $y_j=\delta$ with $y_j\geq\delta$ and $y_j\leq\delta$, respectively. For any two
vectors $\x$ and $\y$ of the same dimension, denote by $\x\succeq \y$ if $x_i\geq y_i$ for all $i$
and by $\x\not\succeq \y$ otherwise.

Consider a fixed $q$-ary matrix $M=[m_{ij}]$ of size $t\times n$. For any vector $\y$ of length $n$,
 we define the {\it syndrome vector} of $\y$ in $M$ by
$\phi_M(\y)=(\phi_1(\y),\phi_2(\y),\cdots,\phi_t(\y))$, where
\begin{equation*}
   \phi_j(\y)= \sum_{i=1}^n y_i\cdot m_{ji}.
\end{equation*} For any two $n$-vectors $\y_0$ and $\y_1$, we say their syndromes $\phi_M(\y_0), \phi_M(\y_1)$ are {\it different},
denoted by $\phi_M(\y_0)\neq \phi_M(\y_1)$, if and only if there exists some $j\in [t]$ such that $\phi_j(\y_0)\neq \phi_j(\y_1)$.
\begin{den}
Let $M=[m_{ij}]$ of size $t\times n$ be a $q$-ary matrix. We say
$M$ is {\it additive $(D,d)$-separable} if $$\phi_M(\y_0)\neq \phi_M(\y_1)$$ for any two $d$-sparse vectors $\y_0, \y_1\in D^n$ with $\y_0\neq \y_1$.
\end{den}
By definition, it is easily seen that additive $(D,d)$-separability is a sufficient and necessary
condition for the considered problem. Moreover, a $q$-ary additive $(D,d)$-separable matrix with
$q=2$ and $D=\{0,1\}$ reduces to a $d$-detecting matrix in \cite{Lindstrom75}. Although
separability provides a solution to identify the unknown vector $\x$, it suffers from lack of
efficient algorithms for decoding.

Disjunct matrices were first studied by Kautz and Singleton \cite{KS64} under the name of
zero-false-drop codes, and also known as cover-free families \cite{EFF85} or superimposed codes
\cite{DR82}. A binary matrix is called $d$-disjunct if it satisfies the  property: for any fixed column and other $d$ columns, there exists a row such that the designated column is 1 and all the $d$ columns are 0. Disjunct matrices have been intensively studied for fifty years. Of particular note is that a $d$-disjunct matrix of size $t\times n$ can identify
up to $d$ defectives with efficient decoding complexity $O(tn)$. Recently, the decoding complexity has been further improved based on other combinatorial structures. Several algorithms with sublinear decoding complexity (in $n$) were proposed \cite{GI04,INR10,NPR11}. It is known \cite{DH00,DR82,DR83} that a $d$-disjunct matrix  of $n$ columns has an upper bound $O(d^2\log n)$ and a lower bound $\Omega(d^2\log n /\log d)$ on the number of rows.  There are many constructions attaining the best known upper bound $O(d^2\log n)$ (see \cite{Dya04, DH06}). Next, we define a new family of disjunct
matrices that can be applied to solve the multi-group testing problem with efficient decoding algorithms.

\begin{den} \label{disjunct}
Let $q,t,n$ and $d$ be positive integers and $w>0$. A $q$-ary matrix $M=[m_{ij}]$ of size $t\times
n$ is called additive $(w,d)$-disjunct if for any $\s\in\{(s_1,s_2,\cdots,s_n)\in \{0,1\}^n:
\sum_{j=1}^ns_j\leq d\}$ and for each $k\in [n]$ such that $s_k=0$, there exists an $i\in [t]$ such
that
$$m_{ik}>w\sum_{j=1}^nm_{ij}s_j.$$
\end{den}

The additive disjunct matrices can be related to some known structures. A conventional $d$-disjunct
matrix is  a binary ($q=2$) additive $(w,d)$-disjunct matrix for any $w\geq 1$. In particular, when
$w\geq q-1$, any $q$-ary additive $(w,d)$-disjunct matrix can be converted simply to a binary
$d$-disjunct matrix by replacing every non-zero entry with 1. As a consequence, we have the
following bound.
\begin{pro}
Let $g(n,w,d,q)$ denote the minimum $t$ such that a $t\times n$ $q$-ary additive $(w,d)$-disjunct
matrix exists. Then $g(n,w,d,q)=O(d^2\log n)$ for any $w\geq 1$ and  $g(n,w,d,q)=\Omega(d^2\log n/\log d)$ when $w\geq q-1$.
\end{pro}

Another is the SQ-disjunct code defined by Emad and Milenkovic \cite{EM12}. When $w=1$, a $q$-ary
additive $(w,d)$-disjunct matrix is reduced to a special case of the
$[q;Q;\eta;(1:d);e]$-SQ-disjunct code with $Q=d$, the thresholds $\eta=[0,1,\cdots,d]^T$ and $e=0$
(error-free). Thus, several useful constructions in \cite{EM13} can be applied immediately for
constructing $q$-ary additive $(w,d)$-disjunct matrices with $w=1$.

Note that in Definition \ref{disjunct} $w$ is assumed only to be positive and needs not to be $w\geq 1$, which is indeed the case
throughout this paper. The case $0<w<1$ seems strange but has its own interest in combinatorial
structure. For instance, when $\frac{1}{2}\leq w<1$ a binary $(w,d)$-disjunct matrix is equivalent
to a matrix satisfying the property that for any fixed column and $d$ other columns there exists a
row such that the designated column has a 1 and the $d$ columns have at least $d-1$ 0's. In view of
this, we believe that the additive disjunct matrices with $0<w<1$ might have other potential
applications.

Next, we study the one-sided case, i.e., elements in $D$ are either all nonnegative or all
nonpositive. By symmetry, we may and shall assume that $D=\{c_0=0,c_1,\cdots,c_m\}$ where
$0<c_1<c_2<\cdots<c_m$.

\begin{thm}
Let $A$ be a $q$-ary additive $(w,d)$-disjunct
matrix of size $t\times n$ with $w=\max_{1\leq k\leq m}\{\frac{c_m-c_{k-1}}{c_k-c_{k-1}}\}$. Then $A$ is additive $(D,d)$-separable.
\end{thm}
\proof Consider any two fixed $d$-sparse vectors $\y, \y'\in D^n$ with $\y\neq \y'$. There exist some $j$'s $\in [n]$ such that $y_j\neq y'_j$. Let $y_j$ be the smallest value among all those $j$'s and by symmetry we may assume that $y_j< y'_j$. Without loss of generality, we assume that $y_j=c_{g-1}$ and therefore $y'_j\geq c_g$. To prove the theorem, it suffices to show that $A\y'\neq A\y$, or equivalently $A\cdot (y'_1,\cdots, y'_j-y_j,\cdots, y'_n)^T \neq A \cdot (y_1,\cdots, y_{j-1}, 0, y_{j+1},\cdots, y_n)^T$. Let $\y_j$ denote the vector $\y$ subject to the $j$-th position, i.e., $(0,\cdots, 0, y_j, 0,\cdots, 0)$. The above inequality can be rewritten as \begin{equation}\label{eq2}A(\y'-\y_j)\neq A(\y-\y_j).\end{equation}

Consider the vector $\s^{\y-\y_j,\geq c_1}$, which is $d$-sparse as is $\y$. Notice that
$s^{\y-\y_j,\geq c_1}_j=0$. By definition of additive $(w,d)$-disjunctness, there exists some
$t^*\in [t]$ such that $a_{t^*j}>w\sum_{i=1}^na_{t^*i}s^{\y-\y_j,\geq c_1}_i$. Since the minimality
of $y_j$, we have $[\y-\y']_i\leq c_m-c_{g-1}$ for all $i$. \begin{align*}
\|[A(\y'-\y_j)-A(\y-\y_j)]_{t^*}\| &\geq (y'_j-y_j)a_{t^*j}-
(c_m-c_{g-1})\sum_{i=1}^na_{t^*i}s^{\y-\y_j,\geq c_1}_i \\
&>\left[(y'_j-y_j)w-(c_m-c_{g-1})\right]\sum_{i=1}^na_{t^*i}s^{\y-\y_j,\geq c_1}_i\\ &\geq
\left[(c_g-c_{g-1})w-(c_m-c_{g-1})\right]\sum_{i=1}^na_{t^*i}s^{\y-\y_j,\geq c_1}_i>0,\end{align*}
where the last inequality holds as $w=\max_{1\leq k\leq m}\{\frac{c_m-c_{k-1}}{c_k-c_{k-1}}\}$.
This proves (\ref{eq2}) and therefore concludes the theorem.\qed

The above theorem makes an attempt to show that the designated matrix $A$ satisfies the
separability property so that the unknown $d$-sparse vector $\x$ can be successfully identified.
However, even ignoring the time for multiplications of vectors, it takes extremely high time
complexity $\d O\left(\sum_{i=0}^d{n\choose i}m^i \right)$ to decode $\x$  by simply applying a
straightforward brute-force procedure based on the separability property. For what follows, we
exploit a more powerful property, disjunctness, of $A$ to quickly identify the unknown vector $\x$.
Next, the focus is on decoding complexity.

\begin{la}\label{basic}
Suppose $\x\in D^n$ is an unknown $d$-sparse vector. Let $A$ be a $q$-ary additive $(w,d)$-disjunct
matrix of size $t\times n$ with $w\geq \frac{c_m}{c_1}$ and let $\v=A\x$ be the outcome vector.
Then $\s^{\x,\geq c_1}$ can be identified from $\v$.
\end{la}
\proof By Lemma \ref{linear}, this problem can be reduced to the problem of learning the unknown
sparse vector $\x'=(x'_1,x'_2,\cdots,x'_n)\in \{0,1,\frac{c_2}{c_1},\cdots,\frac{c_{m}}{c_1}\}^n$
from the outcome vector $\v'=\frac{1}{c_1}\v$.

For each $k$ such that $x'_k\neq 0$, for all $i\in [t]$ we have $\d a_{ik}\leq
\sum_{c=1}^{c_m/c_1}\left(c\sum_{\{j:x'_j=c\}}a_{ij}\right)=v'_i$. Accordingly, $t_k(\v')=0$
whenever $x'_k\neq 0$.

Since the vector $\x$ is $d$-sparse, $\sum_{j=1}^ns^{\x',\geq 1}_j\leq d$. For each $k$ such that $x'_k=0$, by
definition of the $q$-ary additive $(w,d)$-disjunct matrix, there exists $i\in [t]$ such that
$$a_{ik}>w\sum_{j=1}^na_{ij}s^{\x',\geq 1}_j\geq \sum_{j=1}^na_{ij}x'_j=v'_i.$$ This implies that $t_k(\v')>0$ if
$x'_k=0$. By the above discussion, we can identify the vector $\s^{\x',\geq 1}$ through the
counting function $t_k(\v')$. As a consequence, $\s^{\x,\geq c_1}$ can be identified too. \qed

We now analyze the time complexity for the decoding algorithm corresponding to Lemma \ref{basic}.
For each $j\in [n]$, it takes $t$ operations for computing the value $t_j(\cdot)$. Therefore, the
decoding complexity is $O(tn)$.

\begin{cor}\label{ipg}
Let $\x\in\{c_0=0,c_1,\cdots,c_m\}^n$ be an unknown $d$-sparse vector, where
$0<c_1<c_2<\cdots<c_m$. Let $w=\max_{1\leq k\leq m}\{\frac{c_m-c_{k-1}}{c_k-c_{k-1}}\}$. Then any
$q$-ary additive $(w,d)$-disjunct matrix of size $t\times n$ can be used to identify $\x$ with
$O(|D|tn)$ decoding complexity.
\end{cor}
\proof The corollary follows by applying Lemma \ref{basic} repeatedly ($m$ times). The precise
process is as follows. Let $\x^1=\frac{1}{c_1}\x$ and $\v^1=\frac{1}{c_1}\v$ (here we shall use
$\x^i$ and $\v^i$ to denote the updated vectors in the $i$th round). Since $w\geq \frac{c_m}{c_1}$,
by Lemma \ref{basic} we know that $\s^{\x^1,\geq 1}$ can be identified from $\v^1$ successfully.

For $k=2,\cdots,m-1$, define recursively that
$$\v^k=\frac{1}{\frac{c_k-c_{k-2}}{c_{k-1}-c_{k-2}}-1}\left( \v^{k-1}-A\s^{\x^{k-1},\geq
1}\right) \mbox{ and }
\x^k=\frac{1}{\frac{c_k-c_{k-2}}{c_{k-1}-c_{k-2}}-1}\left(\x^{k-1}-\s^{\x^{k-1},\geq 1}\right).$$
It is easily verified that $\v^k=A\x^k$ for all $k$. Note that $\x^k\in
\{0,1,\frac{c_{k+1}-c_{k-1}}{c_k-c_{k-1}},\cdots,\frac{c_{m}-c_{k-1}}{c_k-c_{k-1}}\}^n$ is a
$d$-sparse vector and $w\geq \frac{c_{m}-c_{k-1}}{c_k-c_{k-1}}$ for each $k=1,\cdots,m$. Applying
Lemma \ref{basic} repeatedly, $\s^{\x^k,\geq 1}$ can be identified for all $k=1,\cdots,m$.
Consequently, the unknown vector $\x$ can be identified as $\x=\sum_{i=1}^m
\left(c_i-c_{i-1}\right)\s^{\x^i,\geq 1}=\sum_{i=1}^m c_i\s^{\x,=c_i}.$ This completes the
proof.\qed

\section{The general case}

In this section, we turn our attention to the general case that elements in $D$ are neither all
nonnegative nor all nonpositive. Throughout this section, we shall assume $D=\{z_{m_2},\cdots,z_1,0,c_1,\cdots,c_{m_1}\}$, where
$z_{m_2}<\cdots<z_1<0<c_1<\cdots<c_{m_1}$. For any vector $\v$, $\r\in\{0,1\}^n$ and $h\in\real$,
define
$$\v^{h\r}\equiv \v+hA\r.$$
For any vector $\v$, $d\in \mathbb{N}$ and  $h\in \real$, define
$$t^*_j(\v,h,d)\equiv \min_{\r\in\{0,1\}^n,\|\r\|_0\leq d} t_j(\v^{h\r}) \mbox{ for } j=1,\cdots,n.$$

\begin{thm}\label{basic2}
Let $\x\in D^n$ be an unknown $d$-sparse vector and $w\geq \frac{c_{m_1}-z_{m_2}}{c_1}$. Then
any $q$-ary additive $(w,2d)$-disjunct matrix of size $t\times n$ can be used to identify
$\s^{\x,\geq c_1}$.
\end{thm}
\proof   By Lemma \ref{linear}, this problem is equivalent to the problem of learning $\x\in
\{\frac{z_{m_2}}{c_1},\cdots,\frac{z_1}{c_1},0,1,\frac{c_2}{c_1},\cdots,\frac{c_{m_1}}{c_1}\}^n$.
Let $\x=(x_1,\cdots,x_n)$, $A$ be a $q$-ary additive $(w,2d)$-disjunct matrix, $w\geq
\frac{c_{m_1}-z_{m_2}}{c_1}$, of size $t\times n$ and $\v=A\x$ be the outcome vector corresponding
to $\x$. It suffices to identify $\s^{\x,\geq 1}$.

Consider any vector $\r(\x)=(r_1,r_2,\cdots,r_n)\in \{0,1\}^n$ with $\|\r(\x)\|_0= d$ such that
$\r(\x)\succeq \s^{\x,<0}$. For each $k$ with $x_k\geq 1$, for all $i\in [t]$ we have
\begin{align*}\d v_i^{\frac{-z_{m_2}}{c_1}\r(\x)}&
=v_i-\frac{z_{m_2}}{c_1}\sum_{j=1}^na_{ij}r_j\hspace{3cm}(\mbox{since }\v^{\frac{-z_{m_2}}{c_1}\r(\x)}=\v-\frac{z_{m_2}}{c_1}A\r(\x))\\
&\geq \sum_{\{j:x_j\geq 1\}}a_{ij}+\frac{z_{m_2}}{c_1}\sum_{\{j:x_j<0\}}a_{ij}-\frac{z_{m_2}}{c_1}\sum_{j=1}^na_{ij}r_j\hspace{1cm}(\mbox{since }\r(\x)\succeq \s^{\x,<0})\\
&\geq \sum_{\{j:x_j\geq 1\}}a_{ij}\geq a_{ik}.\end{align*} By definition, $\d
t_k(\v^{-\frac{z_{m_2}}{c_1}\r(\x)})=0$ and therefore $t^*_k(\v,-\frac{z_{m_2}}{c_1},d)=0$ if
$x_k\geq 1$.

Consider the case $x_k< 1$. For any arbitrary vector $\r\in \{0,1\}^n$ with $\|\r\|_0\leq d$,
consider the unknown vector $\s=(s_1,s_2,\cdots, s_n)$ where $$s_\ell= \begin{cases} 1 & \mbox{ if
} x_\ell\geq 1 \mbox{ or }
r_\ell=1,\\
0 & \mbox{otherwise}.
\end{cases}$$ As $\|\r\|_0\leq d$ and $\|\x\|_0\leq d$, we have $\|\s\|_0\leq 2d$.
Since $A$ is a $q$-ary additive $(w,2d)$-disjunct matrix where $w\geq \frac{c_{m_1}-z_{m_2}}{c_1}$,
there exists $i\in [t]$ such that
\begin{align*}\d a_{ik}&>\frac{c_{m_1}-z_{m_2}}{c_1}\sum_{j=1}^na_{ij}s_j&(\mbox{ by definition of disjunctness})&\\
&=\frac{c_{m_1}}{c_1}\sum_{j=1}^na_{ij}s_j-\frac{z_{m_2}}{c_1}\sum_{j=1}^na_{ij}s_j&&\\
&\geq \sum_{j=1}^na_{ij}x_j-\frac{z_{m_2}}{c_1}\sum_{j=1}^na_{ij}s_j&(\mbox{ since $\s \succeq \x$})&\\
&\geq v_i-\frac{z_{m_2}}{c_1}\sum_{j=1}^na_{ij}s_j&&\\
&\geq v_i-\frac{z_{m_2}}{c_1}\sum_{j=1}^na_{ij}r_j&(\mbox{ since $\s \succeq \r(\x)$ and $\frac{z_{m_2}}{c_1}<0$})&\\
&=v_i^{-\frac{z_{m_2}}{c_1}\r}.\end{align*} This implies $\d t_k(\v^{-\frac{z_{m_2}}{c_1}\r})>0$
for any $\r\in \{0,1\}^n$ with $\|\r\|_0\leq d$. Hence, $t^*_k(\v,-\frac{z_{m_2}}{c_1},d)>0$ if
$x_k< 1$.

Therefore, by the above discussion, one can determine whether $x_k\geq 1$ through the counting
function $t^*_k(\v,-\frac{z_{m_2}}{c_1},d)$. \qed

We now analyze the time complexity for the decoding algorithm corresponding to Lemma \ref{basic2}.
For each $j\in [n]$, it takes $t{n\choose d}$ operations for computing the value $t^*_j(\cdot)$.
Therefore, the decoding complexity is $O(tn^{d+1})$.

Combining Corollary \ref{ipg} and Theorem \ref{basic2}, we have the following result.
\begin{thm}\label{general}
Let $\x\in D^n$ be an unknown $d$-sparse
vector and let $$\d w\geq \max\left\{\left\{\frac{c_{m_1}-c_i-z_{m_2}}{c_{i+1}-c_i}\middle\vert 0\leq
i\leq m_1-1\right\}\bigcup \left\{\frac{-z_{m_2}+z_{i-1}}{-z_i+z_{i-1}}\middle\vert 1\leq i\leq m_2
\right\} \right\}.$$ Then any $q$-ary additive $(w,2d)$-disjunct matrix of size $t\times n$ can be
used to identify $\x$ with $O(|D|tn^{d+1})$ decoding complexity.
\end{thm}
\proof By Theorem \ref{basic2}, $\s^{\x,\geq c_1}$ can be identified since $w \geq
\frac{c_{m_1}-z_{m_2}}{c_1}$. Let $\x^1=\x-c_1\s^{\x,\geq c_1}$ and for $i=2,\cdots,m_1-1$ define
recursively $\x^i=\x^{i-1}-(c_{i}-c_{i-1})\s^{\x,\geq c_i}$. Note that $\x^i\in
\{z_{m_2},\cdots,z_1,0,c_{i+1}-c_{i},\cdots,c_{m_1}-c_{i}\}^n$ and $\|\x^i\|_0\leq d$ for
$i=1,\cdots,m_1-1$. With $w\geq \frac{c_{m_1}-c_i-z_{m_2}}{c_{i+1}-c_i}$, applying Theorem
\ref{basic2} repeatedly, one can successfully identify $\s^{\x^i,\geq c_{i+1}-c_i}$ (or
equivalently $\s^{\x,\geq c_{i+1}}$) for $i=1,\cdots,m_1-1$. As a result, one can identify
$\s^{\x,=c_i}=\s^{\x,\geq c_{i}}-\s^{\x,\geq c_{i+1}}$ for all $i=1,\cdots,m_1$ (for
well-defineness let $c_{m_1+1}=\infty$).

Next, let $\d \x'=\left(\sum_{i=1}^{m_1}c_i\s^{\x,=c_i}-\x\right)\in \{0,-z_1,\cdots,-z_{m_2}\}^n$;
hence $\|\x'\|_0\leq d$. As $\d w\geq \max_{1\leq i\leq
m_2}\{\frac{-z_{m_2}+z_{i-1}}{-z_i+z_{i-1}}\}$, by Corollary \ref{ipg}, $\x'$ can be identified.
This completes the proof.\qed

Note that the bound on $w$ in Theorem \ref{general} is not necessary the best choice. One might
obtain a better bound by  first applying the transformation method introduced in Section 2. The
following demonstrates such an example.

\begin{example} Let $\x^1\in \{-2,0,1,4\}^n$ with $\|\x^1\|_0\leq d$ and let $\x^2=-\x^1\in \{-4,-1,0,2\}^n$.
By Lemma \ref{linear}, we know that the problem of learning $\x^1$ is equivalent to the problem
of learning $\x^2$. However, to apply Theorem \ref{general} to $\x^2$, we have to require $w^2\geq
\max\{\frac{2-(-4)}{2-0},\frac{4-0}{1-0},\frac{4-1}{4-1} \}=4$ which is smaller than
$6=\max\{\frac{4-(-2)}{1-0},\frac{4-1-(-2)}{4-1},\frac{2-0}{2-0} \}\leq w^1$, required by simply
applying Theorem \ref{general} to $\x^1$. \qed\end{example}

\section{Constructions for multi-group testing}

This section proposes three constructions for the multi-group testing problem. The first one is to
construct a $q$-ary additive $(w,d)$-disjunct matrix from a conventional binary disjunct matrix by
deleting some rows.

\begin{thm}\label{wd}
Let $A$ be a $q$-ary additive $(w,d)$-disjunct matrix of size $t\times n$ and $w\leq
\frac{q-1}{d}$. If $A$ has $d+2$ rows $R_1,\cdots,R_{d+2}$ pairwise disjoint with entries in
$\{0,1\}$, then $A'$ obtained from $A$ by deleting $R_{d+2}$ and replacing $R_k$ with
$R'_k=(wd+1)R_k+R_{d+2}$ for $k=1,\cdots,d+1$ is a $q$-ary additive $(w,d)$-disjunct matrix of size
$(t-1)\times n$.
\end{thm}
\proof For convenience, represent $A$ using row indices $[t]$ and column indices $[n]$ and without
loss of generality assume $R_1,\cdots,R_{d+2}$ be the first $d+2$ rows, i.e., indexed from 1 to
$d+2$. Let $A'$ be the obtained matrix (using the same indices with $A$). Obviously, every entry in
$A'$ is at most $q$. Consider fixed $j\in [n]$ and $j_1,\cdots,j_d \in [n]\setminus \{j\}$. Since
$A$ is $q$-ary additive
$(w,d)$-disjunct, there exists an $i\in [t]$ such that $a_{ij}\geq w\sum_{k=1}^d a_{ij_{k}}$. There are only three cases as follows.\\
If $i\not\in \{1,\cdots,d+2\}$, then row $i$ is in $A'$ and $a'_{ij}=a_{ij}\geq w\sum_{k=1}^d
a_{ij_{k}}
=w\sum_{k=1}^d a'_{ij_{k}}$, as desired.\\
If $i\in \{1,\cdots,d+1\}$, then row $i$ is in $A'$, $a_{ij}=1$ and $a_{ij_{k}}=0$ for
$k=1\cdots,d$.
Therefore, $a'_{ij}=(wd+1)> w\sum_{k=1}^d 1 \geq w\sum_{k=1}^d  a'_{ij_{k}}$ where the last inequality holds for $a'_{ij_{k}}\leq 1$.   \\
If $i=d+2$, then row $i$ is not in $A'$. We need to find another row $i'$ in $A'$ with the desired
property. In this case, $a_{d+2,j}=1$ and $a_{d+2,j_{k}}=0$ for $k=1\cdots,d$. Since
$R_1,\cdots,R_{d+1}$ are pairwise disjoint, at least one of them, say $i'$, has all $0$ entries at
the columns $j_{k}$'s for $k=1,\cdots,d$, i.e., $a_{i'j_{k}}=0$ for $k=1,\cdots,d$. Since
$a_{d+2,j}=1$ and therefore $a_{i'j}=0$, we have $a'_{i'j}=1$. Thus, $a'_{i'j}>w\sum_{k=1}^d
a'_{i'j_{k}}=0$.

Since $j, j_1,\cdots,j_d$ are chosen arbitrarily, the proof is complete.\qed

A binary matrix $A$ is called {\it transversal} if its rows can be divided into disjoint families
such that rows in each family are disjoint. We say a family of size $b$ if it has $b$ rows. Denote
$f_b(A)$ as the number of disjoint families of size at least $b$ in the matrix $A$.

\begin{example} Let
$$A=\left(
  \begin{array}{cccccccccccccccc}
    1 & 1 & 1 & 1 & 0 & 0 & 0 & 0 & 0 & 0 & 0 & 0 & 0 & 0 & 0 & 0 \\
    0 & 0 & 0 & 0 & 1 & 1 & 1 & 1 & 0 & 0 & 0 & 0 & 0 & 0 & 0 & 0 \\
    0 & 0 & 0 & 0 & 0 & 0 & 0 & 0 & 1 & 1 & 1 & 1 & 0 & 0 & 0 & 0 \\
    0 & 0 & 0 & 0 & 0 & 0 & 0 & 0 & 0 & 0 & 0 & 0 & 1 & 1 & 1 & 1 \\
    \hline
    1 & 0 & 0 & 0 & 1 & 0 & 0 & 0 & 1 & 0 & 0 & 0 & 1 & 0 & 0 & 0 \\
    0 & 1 & 0 & 0 & 0 & 1 & 0 & 0 & 0 & 1 & 0 & 0 & 0 & 1 & 0 & 0 \\
    0 & 0 & 1 & 0 & 0 & 0 & 1 & 0 & 0 & 0 & 1 & 0 & 0 & 0 & 1 & 0 \\
    0 & 0 & 0 & 1 & 0 & 0 & 0 & 1 & 0 & 0 & 0 & 1 & 0 & 0 & 0 & 1 \\
    \hline
    1 & 0 & 0 & 0 & 0 & 1 & 0 & 0 & 0 & 0 & 1 & 0 & 0 & 0 & 0 & 1 \\
    0 & 1 & 0 & 0 & 0 & 0 & 1 & 0 & 0 & 0 & 0 & 1 & 1 & 0 & 0 & 0 \\
    0 & 0 & 1 & 0 & 0 & 0 & 0 & 1 & 1 & 0 & 0 & 0 & 0 & 1 & 0 & 0 \\
    0 & 0 & 0 & 1 & 1 & 0 & 0 & 0 & 0 & 1 & 0 & 0 & 0 & 0 & 1 & 0 \\
  \end{array}
\right)_{12\times 16}.$$ It is easily verified that $A$ is 2-disjunct and transversal as it can be
divided into 3 disjoint families. Further, $f_4(A)=3$.\qed\end{example}

\begin{cor}
Let $A$ be a transversal $d$-disjunct matrix of size $t\times n$. There exists a $q$-ary additive
$(w,d)$-disjunct matrix of size $t'\times n$ with $q\geq wd+2$ and $t'=t-f_{d+2}(A)$.
\end{cor}
\proof The Corollary follows immediately from Theorem \ref{wd}.\qed

In \cite{DHWZ06}, Du et al proved that there exists a transversal $d$-disjunct matrix $A$ of size $t\times n$ with $t=(2+o(1))\left[\frac{d\log n}{\log (d\log n)}\right]^2$ and $f_{d+2}(A)=\frac{d\log n}{\log (d\log n)}$.
As a result, we have the following.
\begin{cor}\label{con1}
Let $q\geq wd+2$. There exists a $q$-ary additive $(w,d)$-disjunct matrix of size  $t\times n$ with $t=(2+o(1))\left[\frac{d\log n}{\log (d\log n)}\right]^2-\frac{d\log n}{\log (d\log n)}.$
\end{cor}

\begin{example} Let
$$A'=\left(
  \begin{array}{cccccccccccccccc}
    5 & 5 & 5 & 5 & 0 & 0 & 0 & 0 & 0 & 0 & 0 & 0 & 1 & 1 & 1 & 1 \\
    0 & 0 & 0 & 0 & 5 & 5 & 5 & 5 & 0 & 0 & 0 & 0 & 1 & 1 & 1 & 1 \\
    0 & 0 & 0 & 0 & 0 & 0 & 0 & 0 & 5 & 5 & 5 & 5 & 1 & 1 & 1 & 1 \\
    5 & 0 & 0 & 1 & 5 & 0 & 0 & 1 & 5 & 0 & 0 & 1 & 5 & 0 & 0 & 1 \\
    0 & 5 & 0 & 1 & 0 & 5 & 0 & 1 & 0 & 5 & 0 & 1 & 0 & 5 & 0 & 1 \\
    0 & 0 & 5 & 1 & 0 & 0 & 5 & 1 & 0 & 0 & 5 & 1 & 0 & 0 & 5 & 1 \\
    5 & 0 & 0 & 1 & 1 & 5 & 0 & 0 & 0 & 1 & 5 & 0 & 0 & 0 & 1 & 5 \\
    0 & 5 & 0 & 1 & 1 & 0 & 5 & 0 & 0 & 1 & 0 & 5 & 5 & 0 & 1 & 0 \\
    0 & 0 & 5 & 1 & 1 & 0 & 0 & 5 & 5 & 1 & 0 & 0 & 0 & 5 & 1 & 0 \\
  \end{array}
\right)_{9\times 16}$$ obtained from $A$ by using operations as in Theorem \ref{wd}. Then $A'$ is a
$6$-ary $(2,2)$-disjunct matrix of size $9\times 16$.\qed\end{example}

 The following two constructions are
based on a special operation of matrices, referred to Kronecker Product. Of particular note
is that the resulting matrices are $q$-ary additive $(D,d)$-separable where $D$ is one-sided. Although the resulting matrices do not satisfy the additive disjunctness property, they also admit efficient decoding algorithms.

\begin{den}
If $A$ and $B$ are matrices of size $n\times m$ and $s\times t$ respectively, then the Kronecker
Product $A\otimes B$ of the two matrices $A$ and $B$ is the $ns\times mt$ matrix whose entries
$(A\otimes B)_{ij,k\ell}=A_{ik}B_{j\ell}$ with row indices listed as
$11,\cdots,1s,21,\cdots,n1,\cdots,ns$ and column indices as $11,\cdots,1t,21,\cdots,m1,\cdots,mt$.
\end{den}
The Kronecker Product of two matrices $A$ and $B$ can also be viewed as $$A\otimes B=\begin{pmatrix} a_{11}B & a_{12}B & \cdots & a_{1m}B \\
a_{21}B & a_{22}B & \cdots & a_{2m}B \\
\vdots & \vdots & \ddots & \vdots \\
a_{n1}B & a_{n2}B & \cdots & a_{nm}B
\end{pmatrix}.$$

\begin{thm}\label{DA}
Let $A$ be a matrix of size $t\times n$ that can successfully identify any unknown $d$-sparse vector $\y\in
\{0,c_1,c_2,\cdots,c_m\}^n$, where $0<c_1<\cdots<c_m$. Let $B$ be a binary
$d$-disjunct matrix of size $t'\times n'$. Then $B\otimes A$ can successfully identify any unknown $d$-sparse
vector $\x\in \{0,c_1,c_2,\cdots,c_m\}^{nn'}$.
\end{thm}
\proof Partition the unknown vector $\x$ equally into $n'$ pieces $\x_1,\x_2,\cdots,\x_{n'}$ where
$\x_j\in \{0,c_1,c_2,\cdots,c_m\}^n$ for all $j=1,\cdots,n'$. Then $$(B\otimes A)\x=\begin{pmatrix} b_{11}A & b_{12}A & \cdots & b_{1n'}A \\
b_{21}A & b_{22}A & \cdots & b_{2n'}A \\
\vdots & \vdots & \ddots & \vdots \\
b_{t'1}A & b_{t'2}A & \cdots & b_{t'n'}A
\end{pmatrix}\begin{pmatrix}\x_1 \\
\x_2\\
\vdots \\
\x_{n'}
 \end{pmatrix}=\begin{pmatrix}\v_1 \\
\v_2\\
\vdots \\
\v_{t'}
 \end{pmatrix},$$ where the outcome vector $\v=\begin{pmatrix}\v_1 \\
\v_2\\
\vdots \\
\v_{t'}
 \end{pmatrix}$ with each $\v_i$ a vector of length $t$.
We now want to show that the unknown vector $\x$ can be identified from the outcome vector $\v$.

We first show that one can determine whether $\x_j=\bold{0}$, where $\bold{0}$ denotes a vector
whose entries are all zero, for all $j$ through the $d$-disjunctness property. Obviously, there are
at most $d$ $\x_j$'s such that $\x_j\neq \bold{0}$ as $\|\x\|_0\leq d$. Let $J$ be such a set
consisting of indices of these $\x_j$'s with $\x_j\neq \bold{0}$. For a fixed $k$ with
$\x_k=\bold{0}$, by the definition of $d$-disjunctness, there exists an $\ell$ such that $b_{\ell
k}=1$ and $b_{\ell j}=0$ for all $j\in J$. It follows that $\v_{\ell}=\begin{pmatrix} b_{\ell 1}A &
b_{\ell 2}A & \cdots & b_{\ell n'}A
\end{pmatrix}\begin{pmatrix}\x_1 \\
\x_2\\
\vdots \\
\x_{n'}
 \end{pmatrix}=\begin{pmatrix}0 \\
0\\
\vdots \\
0
 \end{pmatrix}.$ In contrast, for each $j\in J$, for each $i$ with $b_{ij}=1$ we have $\v_{i}=\begin{pmatrix} b_{i 1}A &
b_{i 2}A & \cdots & b_{i n'}A
\end{pmatrix}\begin{pmatrix}\x_1 \\
\x_2\\
\vdots \\
\x_{n'}
 \end{pmatrix}\neq \begin{pmatrix}0 \\
0\\
\vdots \\
0
 \end{pmatrix}$ since $b_{ij}A\x_j\neq \bold{0}$ and entries in $b_{ij'}A\x_{j'}$ are all nonnegative. As a result, one can determine whether $\x_j$ is
 $\bold{0}$ or not by checking $\v_i=\bold{0}$ for some $i$ with $b_{ij}=1$.

For each $j$ with $\x_j\neq \bold{0}$, by the definition of $d$-disjunctness, there exists an $i$
such that $b_{ij}=1$ and $b_{ij'}=0$ for all $j'\in J\setminus \{j\}$. It follows that
$$\v_{i}=\begin{pmatrix} b_{i 1}A & b_{i 2}A & \cdots & b_{i n'}A
\end{pmatrix}\begin{pmatrix}\x_1 \\
\x_2\\
\vdots \\
\x_{n'}
 \end{pmatrix}=A\x_j.$$ Consequently, $\x_j$ can be identified from the outcome $\v_i$ since $A$ can successfully identify any unknown vector $\y\in
\{0,c_1,c_2,\cdots,c_m\}^n$ with $\|\y\|_0\leq d$, where $0<c_1<\cdots<c_m$. By the above
discussion, one can successfully identify $\x$ by using the matrix $B\otimes A$.\qed

Note that the decoding complexity of Theorem \ref{DA} depends on the decoding complexity of the underlying matrix $A$.

\begin{cor}\label{con2}
Let $\x\in \{0,c_1,c_2,\cdots,c_m\}^{nn'}$ be a $d$-sparse vector, where $0<c_1<\cdots<c_m$. Let $A$
be a $q$-ary additive $(w,d)$-matrix of size $t\times n$ with $w=\max_{1\leq k\leq
m}\{\frac{c_m-c_{k-1}}{c_k-c_{k-1}}\}$ as in Corollary \ref{ipg} and let $B$ be a binary $d$-disjunct matrix of size
$t'\times n'$. Then the $q$-ary matrix $B\otimes A$ can successfully identify the unknown vector
$\x$. Furthermore, the decoding complexity is $O(tt'nn'+dmtn)$.
\end{cor}
\proof The identification result follows immediately from Corollary \ref{ipg} and Theorem \ref{DA}.
The decoding complexity follows by taking $O(tt'nn')$ time in determining $\x_j=\bold{0}$ or not
and then applying the decoding algorithm in Corollary \ref{ipg} to those $\x_j$'s with $\x_j\neq \bold{0}$ at most $d$
times.\qed

Following the idea in \cite{EM12} of concatenating several matrices, we obtain the following
result: Let $A$ be a binary $d$-disjunct matrix of size $t\times n$. Let $u=c_md$ and $n'=\lfloor
\log _u\left(1+(u-1)(q-1)\right)\rfloor$. Construct a $q$-ary matrix $C$ of size $t\times nn'$ by
concatenating $n'$ matrices: $C=\begin{pmatrix} A_1 & A_2 & \cdots & A_{n'}
\end{pmatrix},$ where $\d A_j=\left(\sum _{i=0}^{j-1}u^i\right)A$ for $1\leq j\leq n'$;
equivalently $C=B\otimes A$ where $$B=\begin{pmatrix} 1 & 1+u & 1+u+u^2 & \cdots &
\frac{u^{n'}-1}{u-1}
\end{pmatrix}.$$

\begin{thm}\label{con3}
Let $C$ be the $q$-ary matrix as defined above. Then $C$ can successfully identify any unknown
$d$-sparse vector $\x\in \{0,c_1,\cdots,c_m\}^{nn'}$ with $0<c_1<\cdots<c_m$ where all $c_i$'s are
positive integers. Moreover, the decoding complexity is $O(tnn')$.
\end{thm}
\proof Let $\v=C\x$ be the outcome vector. Consider $\x=\begin{pmatrix} \x_1 & \x_2 & \cdots &
\x_{n'}
\end{pmatrix}$ where $\x_j$'s are vectors of length $n$. We prove this theorem by showing
$\x_{n'},\cdots,\x_1$ can be identified successfully one by one.

Observe that \begin{equation}\label{eq1} \v=\sum_{j=1}^{n'}
A_j\x_j=\left(\sum_{j=1}^{n'-1}\frac{u^j-1}{u-1}A\x_j\right)+\frac{u^{n'}-1}{u-1}A\x_{n'}.
\end{equation}
Since $\d\sum_{j=1}^{n'-1}\|\x_j\|_0\leq d$ and $c_m$ is the largest value, each entry of the term
$\d\left(\sum_{j=1}^{n'-1}\frac{u^j-1}{u-1}A\x_j\right)$ in (\ref{eq1}) is at most
$\d\left(\frac{u^{n'-1}-1}{u-1}\right)c_md <\frac{u^{n'}-1}{u-1}$. It follows that, taking the
floor of numbers in the vector ${\frac{u-1}{u^{n'}-1}}{\v}$ componentwisely, we have $\d\left\lfloor
{\frac{u-1}{u^{n'}-1}}{\v}\right\rfloor =A\x_{n'}$. Note that the above equality holds when $c_i$'s
are positive integers, as required.  This equality implies that $\x_{n'}$ can be identified from the
outcome $\d\left\lfloor {\frac{u-1}{u^{n'}-1}}{\v}\right\rfloor$ as $A$ is a $d$-disjunct matrix,
which guarantees the identification of any $d$-sparse vector with entries all nonnegative in
$O(tn)$ decoding time.

Assume now that $\x_{n'},\x_{n'-1},\cdots,\x_{k+1}$ have been identified. Define
$\v_k=\v-\sum_{j=k+1}^{n'}A_j\x_j$. An analogous argument shows that $\d\left\lfloor
{\frac{u-1}{u^{k}-1}}{\v_k}\right\rfloor =A\x_{k}$. By the $d$-disjunctness property again, one can
identify $\x_k$ from the outcome $\d\left\lfloor {\frac{u-1}{u^{k}-1}}{\v_k}\right\rfloor$.
Repeating this process, $\x_{n'},\cdots,\x_1$ can be identified successfully one by one, and
therefore the total decoding complexity is $O(tnn')$. This completes the proof.\qed

\subsection*{Concluding remarks}

In this section, three constructions are proposed for three purposes: the general case, the one-sided case, and the one-sided integer case. So, it might make little sense to compare their performances in absolute terms.
To conclude this section, we simply list the rate of their performances in comparison with the difficulty of their goals, where the rate of a $t\times n$ matrix $M$ is defined by $\d R(M)=\lim_{n\rightarrow \infty} n/t$. Table 1 lists the corresponding rates of the three constructions. As shown results for the first two constructions, we can conclude that the more restrictions on $D$, the better rates.

{\footnotesize\begin{center}
\begin{threeparttable}
\caption{Rates of three constructions: (1)general case (2)one-sided case (3)one-sided integer case}
\begin{tabular}[t]{lclc}
(1)    &  $\d\frac{(\log (d\log n))^2n}{(d\log n)^2}$  & Cor. \ref{con1}   \\
(2)  &   $\d\frac{(\log (d\log n))^2n}{(d\log n)^2}\times \frac{n}{d^2\log n}$ & Cor. \ref{con2}  \\
(3)     & $\d\frac{n}{d^2\log n}\times (\log q-\log c_md)$ & Thm. \ref{con3}
\end{tabular}
\end{threeparttable}
\end{center}}

Table 2 lists the corresponding decoding complexities for the underlying matrices of the same size $t\times n$.

{\footnotesize\begin{center}
\begin{threeparttable}
\caption{Decoding complexities of three constructions}
\begin{tabular}[t]{lccc}
\hline
 & Corollary \ref{con1} & Corollary \ref{con2} & Theorem \ref{con3}  \\
\hline
general    &  $O(|D|tn^{d+1})$  &  &   \\
one-sided   &  $O(|D|tn)$  &  $O(|D|tn)$ &   \\
one-sided integer   &   &  & $O(tn)$
\end{tabular}
\end{threeparttable}
\end{center}}



\begin{thebibliography}{10}

\bibitem{AS85}
M.~Aigner and M.~Schughart.
\newblock Determining defctives in a linear order.
\newblock {\em J. Statist. Plan. Inform.}, 12:359--368, 1985.

\bibitem{AAS13}
C.~Aksoylar, G.~Atia, and V.~Saligrama.
\newblock Sparse signal processing with linear and non-linear observations: A
  unified shannon theoretic approach.
\newblock In {\em IEEE Information Theory Workshop (ITW2013)}, pages 1--5.
  IEEE, 2013.

\bibitem{AFBC08}
M.~J. Atallah, K.~B. Frikken, M.~Blanton, and Y.~Cho.
\newblock Private combinatorial group testing.
\newblock In {\em 2008 ACM symposium on Information, computer and
  communications security}, pages 312--320, 2008.

\bibitem{AS12}
G.~Atia and V.~Saligrama.
\newblock A mutual information characterization for sparse signal processing.
\newblock In {\em Proc. of Int. Colloq. on Automata, Languages and Programming
  (ICALP)}, 2011.

\bibitem{AS13}
G.~Atia and V.~Saligrama.
\newblock Boolean compressed sensing and noisy group testing.
\newblock {\em IEEE Transactions on Information Theory}, 58(3):1880 -- 1901,
  2012.

\bibitem{BMTW84}
T.~Berger, N.~Mehravari, D.~Towsley, and J.~Wolf.
\newblock Random multipleaccess communication and group testing.
\newblock {\em IEEE Trans. Commun.}, 32:769--779, 1984.

\bibitem{B08}
A.~De Bonis.
\newblock New combinatorial structures with applications to efficient group
  testing with inhibitors.
\newblock {\em J. Combin. Optim.}, 15:77--94, 2008.

\bibitem{BV98}
A.~De Bonis and U.~Vaccaro.
\newblock Improved algorithms for group testing with inhibitors.
\newblock {\em Inform. Process Lett.}, 67:57--64, 1998.

\bibitem{CCH10}
F.~H. Chang, H.~Chang, and F.~K. Hwang.
\newblock Pooling designs for clone library screening in the inhibitor complex
  model.
\newblock {\em J. Comb. Optim.}, 22:145--152, 2011.

\bibitem{CCF10}
H.~L. Chang, H.~B. Chen, and H.~L. Fu.
\newblock Identification and classification problems on pooling designs for
  inhibitor models.
\newblock {\em J. Comput. Biol.}, 17(7):927--941, 2010.

\bibitem{CB11}
H.B. Chen and A.~De Bonis.
\newblock An almost optimal algorithm for generalized threshold group testing
  with inhibitors.
\newblock {\em J. Comput. Biol.}, 18:851--864, 2011.

\bibitem{CF13}
H.B. Chen and H.~L. Fu.
\newblock Group testing with multiple mutually-obscuring positives.
\newblock {\em LNCS: Information Theory, Combinatorics, and Search Theory},
  7777:557--568, 2013.

\bibitem{CKW06}
J.~Cheng, K.~Kamoi, and Y.~Watanabe.
\newblock Spreading set with error correction for multiple-access adder
  channel.
\newblock {\em IEEE Trans. Inform. Theory}, 52(12):5524--5529, 2006.

\bibitem{CW99}
J.~Cheng and Y.~Watanabe.
\newblock T-user code with arbitrary code length for multiple-access adder
  channel.
\newblock In {\em IEICE Trans. Fundamentals}, volume E82-A, pages 2011--2016,
  1999.

\bibitem{CM11}
M.~Cheng and Y.~Miao.
\newblock On anti-collusion codes and detection algorithms for multimedia
  fingerprinting.
\newblock {\em IEEE Trans. Inform. Theory}, 57(7):4843--4851, 2011.

\bibitem{Dam98}
P.~Damaschke.
\newblock Randomized group testing for mutually obscuring defectives.
\newblock {\em Inf. Process. Lett.}, 67:131--135, 1998.

\bibitem{Dam05}
P.~Damaschke.
\newblock Threshold group testing.
\newblock In {\em General Theory of Information Transfer and Combinatorics:
  LNCS}, volume 4123, pages 707--718, 2005.

\bibitem{DD11}
A.~De~Bonis and G~Di~Crescenco.
\newblock Combinatorial group testing for corruption localizing hashing.
\newblock In B.~Fu and D.Z. Du, editors, {\em LNCS}, volume 6842, pages
  579--591. Springer, 2011.

\bibitem{DH00}
D.~Z. Du and F.~K. Hwang.
\newblock {\em Combinatorial Group Testing and Its Applications}.
\newblock World Scientific, $2nd$ ed. edition, 2000.

\bibitem{DH06}
D.~Z. Du and F.~K. Hwang.
\newblock {\em Pooling Designs and Nonadaptive Group Testing - Important Tools
  for DNA Sequencing}.
\newblock World Scientific, 2006.

\bibitem{DHWZ06}
D.Z. Du, F.~K. Hwang, W.~Wu, and T.~Znati.
\newblock New construction for transversal design.
\newblock {\em Journal of Computational Biology}, 13:990--995, 2006.

\bibitem{DMTV01}
A.~G. D'yachkov, A.~J. Macula, D.~C. Torney, and P.~A. Vilenkin.
\newblock Two models of nonadaptive group testing for designing screening
  experiments.
\newblock In {\em Proc. 6th Inter. Workshop in Model Oriented Design and
  Analysis}, pages 63--75. Physica-Verlog, 2001.

\bibitem{DR82}
A.~G. D'yachkov and V.~V. Rykov.
\newblock Bounds on the length of disjunct codes.
\newblock {\em Problemy Peredachi Inform.}, 18(3):7--13, 1982.

\bibitem{DR83}
A.~G. D'yachkov and V.~V. Rykov.
\newblock A survey of superimposed code theory.
\newblock {\em Problems Control Inform. Theory}, 12:229--242, 1983.

\bibitem{Dya04}
A.G. D'yachkov.
\newblock Lectures on designing screening experiments.
\newblock In {\em Lecture Note Series 10}, pages (monograph, pp.~112).
  Combinatorial and Computational Mathematics Center, Pohang University of
  Science and Technology (POSTECH), Korea Republic, 2004.

\bibitem{EM12}
A.~Emad and O.~Milenkovic.
\newblock Semi-quantitative group testing.
\newblock In {\em Proc. IEEE Int. Symp. Inf. Theory (ISIT12)}, pages
  1847--1851, July 2012.

\bibitem{EM13}
A.~Emad and O.~Milenkovic.
\newblock Semiquantitative group testing.
\newblock {\em IEEE Trans. Inf. Theory}, 2014.
\newblock doi: 10.1109/TIT.2014.2327630.

\bibitem{EFF85}
P.~Erd\H{o}s, P.~Frankl, and Z.~F\H{u}redi.
\newblock Family of finite sets in which no set is coverd by the union of $n$
  others.
\newblock {\em Israel J. Math.}, 51:79--89, 1985.

\bibitem{FKKM97}
M.~Farach, S.~Kannan, E.~Knill, and S.~Muthukrishnan.
\newblock Group testing problem with sequences in experimental molecular
  biology.
\newblock In {\em Proc. Compression and Complexity of Sequences}, pages
  357--367, 1997.

\bibitem{GIS08}
A.~C. Gilbert, M.~A. Iwen, and M.~J. Strauss.
\newblock Group testing and sparse signal recovery.
\newblock In {\em The 42nd Asilomar Conference on Signals, Systems and
  Computers}, pages 1059--1063, 2008.

\bibitem{GI04}
V.~Guruswami and P.~Indyk.
\newblock Linear-time list decoding in error-free settings.
\newblock In {\em Proceedings of of the 31st International Colloquium on
  Automata, Languages and Programming (ICALP)}, pages 695--707, 2004.

\bibitem{HC07}
F.~K. Hwang and F.~H. Chang.
\newblock The identification of positive clones in a general inhibitor model.
\newblock {\em J. Comput. System Sci.}, 73:1090--1094, 2007.

\bibitem{INR10}
P.~Indyk, H.~Q. Ngo, and A.~Rudra.
\newblock Efficiently decodable non-adaptive group testing.
\newblock In {\em The 21st Annual ACM-SIAM Symposium on Discrete Algorithms
  (SODA 10)}, 2010.

\bibitem{Jevtic}
D.~B. Jevti\'{c}.
\newblock On families of sets of integral vectors whose representatives form
  sum-distinct sets.
\newblock {\em SIAM J. DISC. MATH.}, 8:652--660, 1995.

\bibitem{KS64}
W.~H. Kautz and R.~R. Singleton.
\newblock Nonrandom binary superimposed codes.
\newblock {\em IEEE Trans. Inform. Theory}, 10:363--377, 1964.

\bibitem{MT11}
M.~M\'{e}zard and C.~Toninelli.
\newblock Group testing with random pools: optimal two-stage algorithms.
\newblock {\em IEEE Trans. Inform. Theory}, 57:1736--1745, 2011.

\bibitem{PS94}
R.~M. Phatarfod and A.~Sudbury.
\newblock The use of a square array scheme in blood testing.
\newblock {\em Stat. Med.}, pages 2337--2343, 1994.

\bibitem{XTSY01}
M.~Xie, K.~Tatsuoka, J.~Sacks, and S.~Young.
\newblock Group testing with blockers and synergism.
\newblock {\em J. Amer. Stat. Assoc.}, pages 92--102, 2001.

\bibitem{XSTZ10}
Y.~Xuan, I.~Shin, M.T. Thai, and T.~Znati.
\newblock Detecting application denial-of-service attacks: a
  group-testing-based approach.
\newblock {\em IEEE Trans. Par. Distr. Syst.}, 21:1203--1216, 2010.

\end{thebibliography}
\end{document}